\documentclass[conference]{IEEEtran}
\usepackage[T1]{fontenc}
\usepackage{gensymb}
\usepackage{todonotes}
\usepackage{booktabs}
\usepackage{multirow}
\usepackage{amsmath}
\usepackage{afterpage}
\usepackage[noadjust]{cite}
\usepackage{graphicx}
\usepackage{subcaption}
\usepackage{enumitem}
\usepackage{dsfont}

\DeclareMathOperator*{\argmax}{argmax}
\ifCLASSINFOpdf
\else
\fi
\usepackage{url}

\begin{document}
%
\title{Capturing Capacity and Profit Gains with Base Station Sharing in mmWave Cellular Networks}

\author{\IEEEauthorblockN{Shahram~Shahsavari, Fraida~Fund, Elza~Erkip, Shivendra~S.~Panwar}
\IEEEauthorblockA{Department of Electrical and Computer Engineering\\
NYU Tandon School of Engineering\\
\{\url{shahram.shahsavari, ffund, elza, panwar}\}@nyu.edu}
}


%


\maketitle

\begin{abstract}
Due to the greater path loss, shadowing, and increased effect of blockage in millimeter
wave cellular networks, base station sharing among network 
service providers has the potential to significantly improve overall 
network capacity. However, a service provider may find that 
despite the technical gains, sharing actually reduces its profits because
it makes price competition between service providers tougher.
In this work, a weighted scheduling algorithm is described, which gives greater control
over how the airtime resource is allocated within a shared cell. 
It is shown that, under certain market conditions, there exist 
scheduling weights such that base station sharing is more profitable 
than not sharing for both service providers in a duopoly market, 
while still achieving almost as much network capacity 
as in a conventional base station sharing scenario. Thus, 
both technical and economic benefits of base station 
sharing may be realized simultaneously.
\end{abstract}


%
\IEEEpeerreviewmaketitle

\section{Introduction}

Compared to conventional microwave frequencies, millimeter wave (mmWave) 
frequencies have several properties that suggest greater 
technical benefits due to spectrum and base station sharing
among cellular network service providers.
The massive bandwidth available at mmWave bands is unlikely to be 
fully utilized by any one service provider, and the narrow beamwidth of mmWave
signals decreases the likelihood of interference due to 
uncoordinated spectrum sharing. Meanwhile, the greater path 
loss and increased spatial degrees of freedom necessitate 
a dense deployment of base stations, which may be costly
for a single service provider.

Early results in the literature suggest that in fact, 
mmWave cellular networks can benefit on a technical level from resource 
sharing~\cite{andrews-sharing,matia,european-sharing,andrews-sharing-second}, 
Under certain conditions (e.g. sufficiently narrow beamwidth and low enough 
base station density)
the network capacity and the data rate experienced
by most users are higher when service providers share base stations
and spectrum, even without coordination.
Some have suggested that these technical gains may
translate to economic gains. For example,~\cite{andrews-sharing}
claims that it is economical for mmWave service providers to share resources
because they can offer the same quality of service while
licensing less spectrum, while~\cite{matia} points out savings 
on costs including both spectrum licensing and base station deployment. 
Similarly,~\cite{andrews-sharing-second} suggests that a 
mmWave spectrum holder can earn additional revenue by
licensing the spectrum in a secondary market with the condition of
restricted interference to its own users.

However, even if service providers can reduce costs or earn
revenue from secondary licensing while keeping quality of
service the same, resource sharing can affect profits if it shifts
demand to a competing service provider, or if it changes the
market dynamics in a way that forces down the price. 
Our previous work~\cite{fund2017resource} suggests that 
mmWave cellular service providers may be \emph{less} likely
to consider sharing resources in a competitive market. 
In mmWave cellular networks, a service provider with 
a large network (i.e. having more spectrum and base stations)
has a considerable advantage over a smaller service provider - 
much more so than in an equivalent microwave network.
With resource sharing, the large service provider would
lose this considerable competitive advantage and would 
have to deal with stiffer price competition, and so 
would be unwilling to consider resource sharing 
under certain market conditions.
Furthermore, a service provider with a large network
typically finds that it stands to gain much less from pooling 
resources with a small service provider, than the small 
provider does. Thus, we find that with standard base station 
sharing, it is difficult to capitalize on both technical and
profit gains at the same time.

To address this issue, we turn to cooperative game theory, where
we find several methods for distributing 
the profits of a coalition in a weighted manner 
among its members. These methods have also been used to allocate network 
or computational resources in various 
settings~\cite{singhProfitSharing,Iturralde2013,allocateFairPayoff,eyeball}. 
In this work, 
we explore the feasibility of using a similar approach 
to allocate airtime within a wireless cell among members
of a coalition of network service providers that share
base stations, with the goal of finding a resource allocation 
scheme under which both technical and profit gains
may be realized.


The contributions of this work are as follows:

\begin{itemize}
  \item We describe a scheduling approach that allows for control over how 
  airtime is allocated among users in a shared cell, so that it can be weighted according 
  to the relative contributions (in terms of base stations) 
  of the network service providers
  to which they subscribe. We show that with this scheduling 
  approach we can achieve almost all of the efficiency gains associated
  with resource sharing, while still preserving a difference in quality 
  of service between service providers.
  \item We formulate a simple resource sharing game, and 
  show that in a competitive duopoly market for mmWave cellular service, 
  network service providers may be willing to share base station 
  resources with our proposed scheduler.
\end{itemize}

The rest of this paper is organized as follows. 
First, in Section~\ref{sec:scheduler}, we propose 
an approach to scheduling that weighs users of different
network service providers according to their contributions
to the pool of shared resources. Next, in Section~\ref{sec:simulation}, 
we show through simulations that with this scheduler we can achieve almost
the full technical sharing gains (i.e. overall network capacity) 
while still distinguishing between 
network service providers in terms of the quality of service 
offered to subscribers. In Section~\ref{sec:game}, we model a 
simple duopoly market as a game, and show that, under certain 
conditions, both service providers in the market earn 
higher revenue while sharing resources using the weighted scheduler.
Finally, in Section~\ref{sec:conclusion} we conclude with discussion
and suggestions for future work.

\section{Scheduler for weighted resource sharing}
\label{sec:scheduler}

We consider a system with a set of wireless base stations (BSs), users, 
and NSPs. Each BS is operated by one network service provider (NSP), and each 
user subscribes to one NSP. When there is no base station sharing, each 
user is served by the closest BS operated by the NSP to which it subscribes.
However, when NSPs form a coalition to share base stations, then their subscribers
can be served by the closest BS that is operated by an NSP that is a member 
of the coalition. All BSs use the same unlicensed spectrum, regardless
of whether or not they are shared.

In this section, we propose a scheduler for allocating downlink time slots
in a shared wireless cell serving users of multiple network 
service providers, that satisfies the following
design goals:

\begin{itemize}
  \item The scheduler is opportunistic, so that it can take advantage of 
  the multi-user diversity that is a major factor
  in the sharing gains observed in mmWave cellular networks.
  \item It is temporal fair, so that in the long term, the airtime
  allocated to each user will converge to a predefined share of the 
  total airtime.
  \item By setting the predefined shares, we 
  can differentiate between subscribers of multiple NSPs 
  so that a large NSP can maintain some competitive advantage.
\end{itemize}

We adopt a modified scheduler based on
the multicell temporal fair opportunistic scheduler proposed
in~\cite{shahram-scheduler}. 
At each time slot, the scheduler selects a user $j^*=\argmax_j(R_j+\gamma b_j)$, where $R_j$ denotes the estimated data rate of the user $j$ (using a pilot sequence) 
and $b_j$ is a credit parameter updated as $\forall j: b_j=b_j+a_j-\mathds{1}_{\{j=j^*\}}$ to achieve long-term temporal fairness among the users. 
This scheduler guarantees that the temporal share of user $j$ (i.e. the fraction of the time slots in which user $j$ is chosen) converges to a predefined weight $a_j$ ($\sum_{j} a_j = 1$) 
while exploiting the multi-user diversity to increase the
total throughput in the cell~\cite{shahram-scheduler}. 
In other words, this scheduler can be viewed as an opportunistic 
fair credit-based procedure where selected (not selected) 
users lose (gain) credit and the
algorithm parameter $\gamma \geq 0$ is the weight of the credit
component. 
For very large $\gamma$, this scheduler is equivalent to a round robin scheduler
because the users are chosen almost only based on their credit parameter. 
On the other hand, for $\gamma=0$ the scheduler is purely opportunistic and does 
not offer temporal fairness. 

The operation of the scheduler depends on the base station sharing scenario:
\begin{itemize}
  \item \textbf{No base station sharing}: all users in a cell 
  are assigned the same weight $a_j = \frac{1}{N_i}$, 
  where $N_i$ is the number of subscribers in the cell. All
  users in the cell subscribe to NSP $i$, since with no BS sharing, 
  users are only served by a base station operated by the NSP to which they subscribe.
  A user is not necessarily served by the closest base station.
  \item \textbf{Equal sharing}: BSs are shared by a set of NSPs $I$.
  In each shared cell, all users in the cell are assigned the 
  same weight $a_j = \frac{1}{N_I}$, where $N_I$ is the number of subscribers
  (of any NSP in $I$) in the cell.
  \item \textbf{Weighted sharing}: BSs are shared by a set of NSPs $I$.
  In each shared cell, users are assigned weights
  $a_j = \frac{\psi_i}{N_i}$, where $\psi_i$ is the weight assigned to NSP $i\in I$, $\sum_{i \in I} \psi_i = 1$ 
  and user $j$ subscribes to NSP $i$. 
\end{itemize}
Note that the total number of BSs in the system, 
and the total number of users, is fixed, but cell boundaries are
different depending on whether BSs are shared or not. When BSs are shared, 
users may be served by a closer BS, so the average coverage area of a 
cell will decrease while the average number of users served remains the same.

In the weighted sharing scenario, 
airtime is not equally shared among all users in a cell, 
but is allocated based on a per-NSP parameter $\psi_i$. 
This parameter can be adjusted in order to 
differentiate between subscribers of different NSPs.
In Section~\ref{sec:game}, we will show 
that for certain values of $\psi_i$, weighted sharing may be mutually
beneficial for NSPs in a duopoly market.

\section{Network simulation}
\label{sec:simulation}


In this section, we show by means of simulation that
in the weighted sharing scenario, we can 
achieve almost the same sharing gains observed 
for the equal sharing scenario in a mmWave cellular network.
First, we describe the system model underlying the simulations. 
Then, we show simulation results for a network 
with two symmetric NSPs (same density of base stations 
and subscribers) and several networks with two asymmetric NSPs.

\subsection{Technical system model}

Our simulation captures the following key characteristics
of mmWave networks:
\begin{itemize}
\item {\textbf{Channel model}: We use the empirically derived
line of sight (LOS), NLOS and outage
probabilistic channel models for mmWave links
from~\cite{mustafa-channel}.}
\item {\textbf{Directional transmission}: We use the antenna pattern model
described in \cite{heath-mm} with model parameters representing
an 8x8 antenna array at the BS and a 4x4 antenna array at the user.}
\end{itemize}

We consider a system with two NSPs, with a 
system model similar to~\cite{fund2017resource}. 
Each NSP $i \in\{1,2\}$
has BSs distributed in the network area using a
homogeneous Poisson Point Process (hPPP) with intensity $\lambda^B_i$,
and users whose locations are modeled by an
independent hPPP with intensity $\lambda^U_i$.
Also, both NSPs use the same frequency band with bandwidth $W$.
Although it is possible to have strong interference
due to the shared frequency with no coordination,
the narrow beamwidth, increased channel loss,
and large bandwidth (hence large noise power) in mmWave
networks means that noise and not interference is
usually the dominant effect \cite{mustafa-channel}.

\begin{figure}
    \centering
\includegraphics[width=0.45\textwidth]{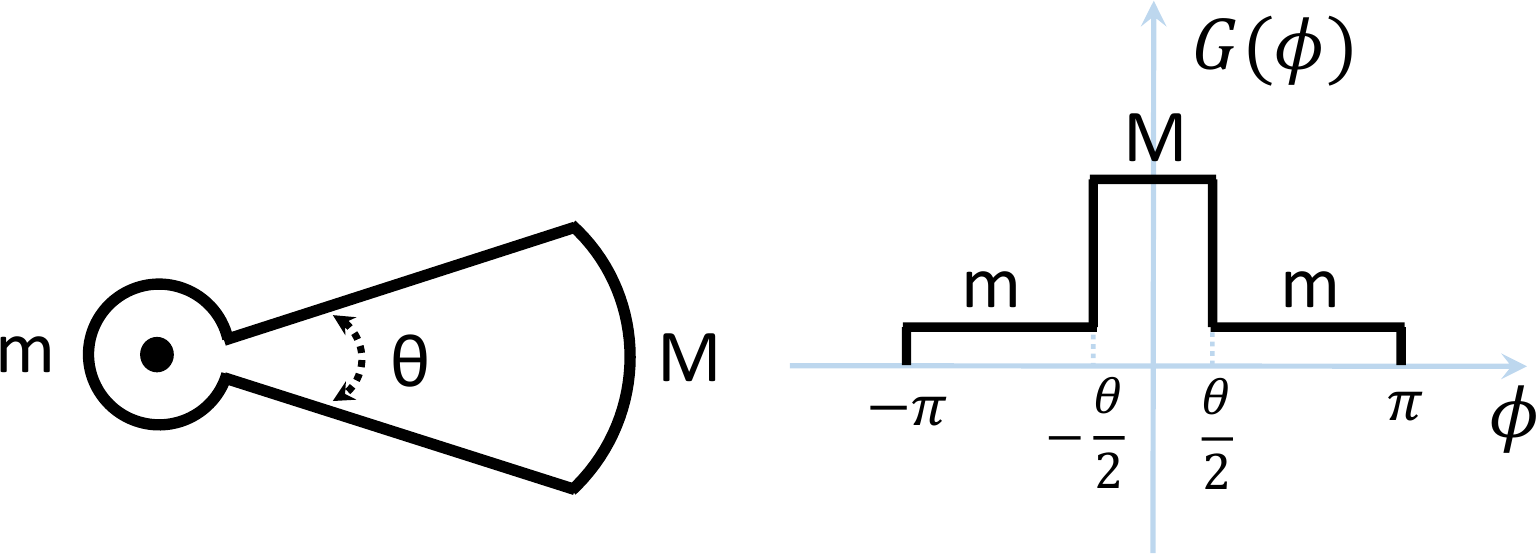}
\caption[]{Simplified antenna pattern with main lobe $M$, back lobe $m$ and beamwidth $\theta$.}
 \label{pattern}
\end{figure}

Both BSs and users use antenna arrays for directional beamforming.
We approximate the actual array patterns
using a simplified pattern as in \cite{heath-mm}.
Let $G(\phi)$ denote the
antenna directivity pattern depicted in Fig. \ref{pattern},
where $M$ is the main lobe power gain, $m$ is the back lobe
gain and $\theta$ is the beamwidth of the main lobe.
In general, $m$ and $M$ are dependent on 
the number of antennas in the array and $M/m$ depends
on the type of the array. Furthermore,
$\theta$ is inversely proportional to the number of
antennas, i.e., the greater the number of antennas,
the more beam directionality. We let $G^B(\phi)$
(which is parameterized by $M^B$, $m^B$, and $\theta^B$) be
the antenna pattern of the BS, and $G^U(\phi)$
(which is parameterized by $M^U$, $m^U$, and $\theta^U$) be the antenna pattern
of the user.

We model a time-slotted downlink of a cellular system.
For path loss, shadowing,
and outage, line of sight (LOS), and NLOS probability distributions,
we use models adopted from \cite{mustafa-channel}.
We assume Rayleigh block fading to model small scale channel variations. The data rate is modeled as
\begin{equation}
R=(1-\alpha)W\log_2 \Bigg(1+\beta \frac{PG^U(0)G^B(0)H}{N_fN_0W+Y}\Bigg),
\label{rate-model}
\end{equation}
where $\alpha$ and $\beta$
are overhead and loss factors, respectively,
and are introduced to fit a specific physical layer to the
Shannon capacity curve \cite{mustafa-channel}.
Furthermore, $P$ is the BS transmitting power,
and $H$ is the channel power gain derived from the model
discussed above, incorporating the effects of fading, shadowing, outage, and path loss.
We assume perfect beam alignment between BS and user device within a cell,
therefore the antenna power gain (link directionality) is
$G^U(0)G^B(0)=M^UM^B$.
Finally, $N_f$, $N_0$, $W$ and $Y$ are user device noise figure,
noise power spectral density, bandwidth, and interference power, respectively.

\subsection{Results}

Using the model described in the previous subsection, 
we simulate a mmWave network with the parameters given in
Table~\ref{sim-param}.

\begin{table}[h]
\caption{Network parameters}
\centering
\begin{tabular}{ll}
\toprule \textbf{Parameter} & \textbf{Value}   \\
\midrule
Frequency                         &  73 GHz          \\
Bandwidth ($W$)                   &  1 GHz           \\
Total BS density $\big(\lambda^B = \sum_{i \in \{1,2\}}\lambda^B_{i}\big)$      &  100 BSs/$\text{km}^2$   \\
Total user density $\big( \lambda^U = \sum_{i \in \{1,2\}}\lambda^U_{i}\big)$    &  500 UEs/$\text{km}^2$   \\
BS transmit power $P$             &  30 dBm                   \\
$\big(M^B$, $m^B$, $\theta^B\big)$          & (20 dB, -10 dB, 5\degree)  \\
$\big(M^U$, $m^U$, $\theta^U\big)$          & (10 dB, -10 dB, 30\degree) \\
Simulation area                              & 1 $\text{km}^2$  \\
Rate model ($\alpha$, $\beta$)    & (0.2, 0.5)     \\
User device noise figure ($N_f$)             & 7 dB        \\
Noise PSD ($N_0$)                   & -174 dBm/Hz     \\
Scheduler parameter ($\gamma$)                   & 0.01     \\

\bottomrule
\end{tabular}
\label{sim-param}
\end{table}

We simulate a network with two NSPs, and a fixed number 
of BSs and users
divided among the NSPs according to their relative size $n_i$.
An NSP $i$ has BS density $\lambda^B_{i} =  n_i \lambda^B$ and 
user density $\lambda^U_{i} = n_i \lambda^U$.
We consider the case of symmetric NSPs ($n_1 = n_2 = 0.5$)
and two cases of asymmetric NSPs 
($n_1 = 0.6, n_2 =  0.4$ and $n_1 = 0.7, n_2=0.3$).

\begin{figure*}
    \centering
     \begin{subfigure}[b]{\textwidth}
        \includegraphics[width=\textwidth]{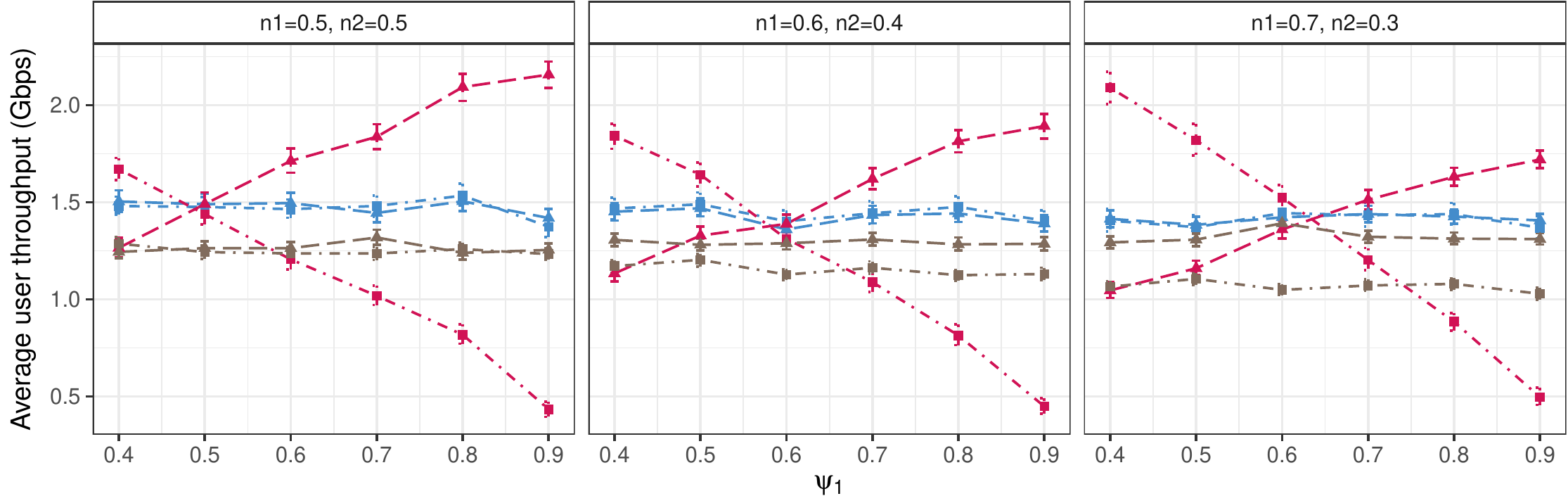}
        \caption{Average user throughput.    \vspace{2em}}
       \label{fig:user-tput}
    \end{subfigure}
    \begin{subfigure}[b]{\textwidth}
      \includegraphics[width=\textwidth]{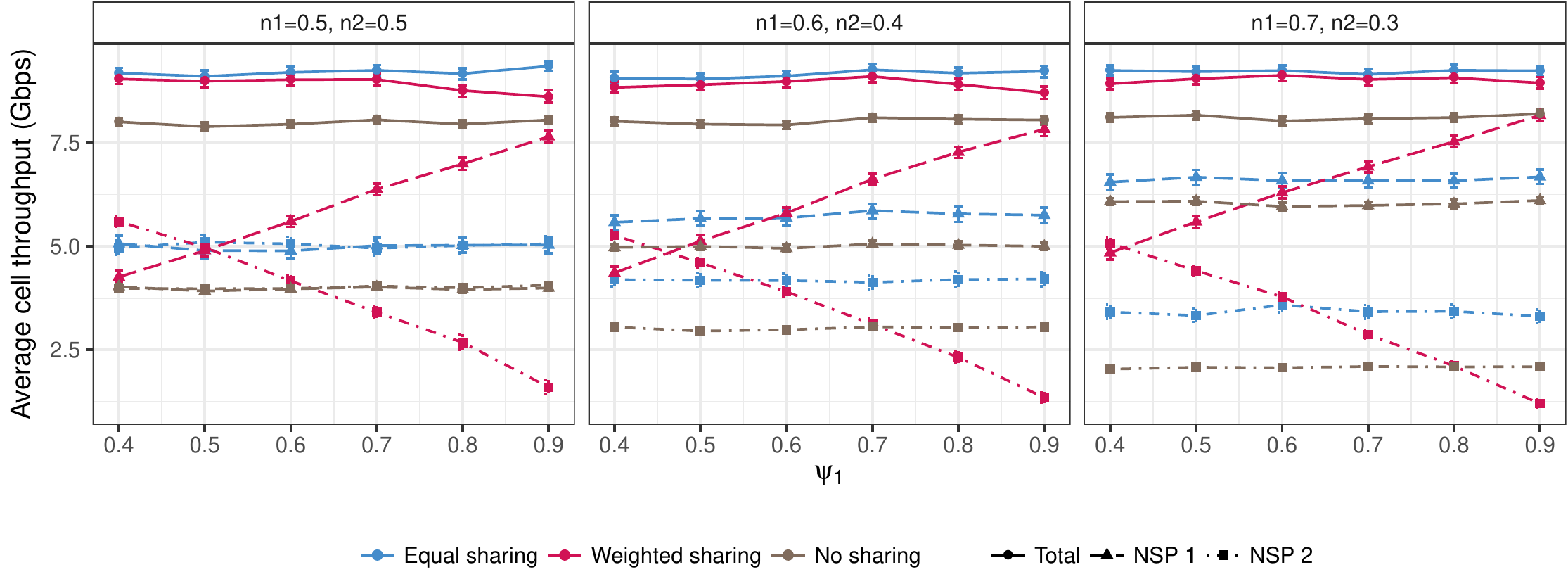}
      \caption{Average cell throughput.}
      \label{fig:cell-tput}
    \end{subfigure}
            \caption[]{Average user and cell throughput in three sharing scenarios
             - no sharing, equal sharing, and weighted sharing -
      for symmetric and asymmetric NSPs. The share of the total number of BSs belonging to 
      NSP~1 and 2, $n_1$ and $n_2$ respectively, are shown at the top of each panel.
      In the lower plot, we also show the total average throughput achieved by NSP~1 and NSP~2 together
      within one cell, in addition to the individual throughputs measured by each NSP
      within the cell.
      The horizontal axis shows how throughput varies with $\psi_1$, the weight
      assigned to (the larger) NSP~1 in the scheduler.
      Error bars show 95\% confidence intervals. For the scenarios with equal sharing or no sharing,
      the throughput is flat with only minor variation within the confidence intervals. }
    \label{fig:sim-results}
\end{figure*}

The simulation results are shown in Figure~\ref{fig:sim-results}.
Figure~\ref{fig:user-tput} shows the average user throughput.
Figure~\ref{fig:cell-tput} shows the 
throughput for each NSP separately,   
as well as their sum. 

The average user throughput
is higher with equal sharing than with no sharing for 
users of both NSPs (Figure~\ref{fig:user-tput}).
However, even in an asymmetric scenario, where one
NSP contributes more base stations to the pool of shared
resources than the other, users of both NSPs can expect exactly the same
average throughput. The larger NSP loses any competitive
edge it had in the market, which disincentivizes base station sharing
despite the higher average data rate experienced by its users. 
With weighted sharing, however,
$\psi_1$ (and consequently, $\psi_2$) 
can be adjusted so as to preserve a distinction
in the quality of service offered to users of different NSPs.
In Section~\ref{sec:game} we will show that for some market scenarios, 
there are values of $\psi_1$ and $\psi_2$ with which weighted sharing
is mutually beneficial for both NSPs.

Meanwhile, considering the average cell 
throughput (Figure~\ref{fig:cell-tput}),
we note that in the weighted sharing scenario the 
total throughput achieved by NSP~1 and NSP~2 
together, is almost as high as in the equal sharing 
scenario. The equal sharing scheduler is slightly 
more opportunistic than the weighted sharing scheduler, 
leading to higher overall throughput.
However, in the weighted 
sharing scenario, the network still benefits from 
greater BS diversity and multiuser diversity
than when there is no sharing at all, capturing 
most of the potential sharing gains.

\section{Simple duopoly game}
\label{sec:game}

We have shown in Section~\ref{sec:simulation} that with the weighted sharing 
scheduler, mmWave NSPs can benefit from technical sharing gains 
(achieve an overall network capacity similar to equal sharing) while 
still preserving a competitive difference in user quality of service. 
This has the potential to create market dynamics that are more favorable to resource sharing. 
In this section, we investigate a simple duopoly market and show that 
under some market conditions where equal sharing is not mutually beneficial 
to both NSPs, weighted resource sharing may be. In those markets
where weighted sharing can be mutually beneficial, we also give the range of $\psi_1$ and
$\psi_2$ for which both NSPs earn higher profits with weighted sharing than with none.

\subsection{Scenario}


We consider a simple duopoly game involving three players: a set of consumers, 
a dominant NSP~1 with size $n_1$, and a smaller NSP~2 with size $n_2 < n_1$
where $n_i$ represents the share of base stations operated by NSP $i$.
The game is played in three stages:
\begin{enumerate}
\item NSP~1 sets the price of its service, $p_1$.
\item NSP~2 sets the price of its service, $p_2$.
\item Each consumer subscribes to one NSP or to neither.
\end{enumerate}

Each NSP $i \in \{1,2\}$ seeks to maximize its profits
\begin{equation}
\pi_i(p_i) = p_i N_i - c_i N_i
\label{eq:nspprofit}
\end{equation}
where $N_i$ is the number of subscribers of NSP $i$ and will be determined
by the decisions of the consumers' decisions in the last stage of the 
game, and $c_i$ is the marginal cost
to the NSP of serving one subscriber. An NSP can increase its profits by raising 
the price of service, but this will affect the number of subscribers it captures.

Consumers trying to maximize their individual surplus 
evaluate the competing services in terms of the difference in size
$n_i$, with a larger network representing more base stations and consequently better service, 
as well as in terms of the difference in price $p_i$. 
We have heterogeneous consumers parameterized by taste parameter $\omega$,
which represents the degree to which the consumer values the wireless service,
with $\omega$ distributed uniformly from $[0, \hat{\omega}]$.
The surplus of a consumer of type $\omega$ subscribing 
to NSP $i$ depends on the resource sharing scenario, the network configuration, and 
the relative share of base stations operated by its NSP.
We use the average data rate as a metric of utility, 
and for mmWave networks we model it as a linear function of $n_i$ (which we confirm empirically 
with the simulation in Section~\ref{sec:simulation}), 
with a parameter $\mu$ capturing 
the network configuration.

When there is no resource sharing, the surplus of a consumer of type $\omega$ subscribing 
to NSP $i$ is
\begin{equation}
 u(\omega, n_i, p_i ) =  \omega \mu n_i - p_i
\end{equation}
\noindent for $i \in \{1,2\}$,
and a consumer subscribes to at most one NSP.

If the NSPs share their mmWave network resources,
then the surplus of a consumer of type $\omega$ subscribing 
to NSP $i$ is
\begin{equation}
 u\Big(\omega, \sum_{k \in \{1,2\}} n_k, p_i \Big) = \omega \mu \sum_{k \in \{1,2\}} n_k - p_i
\end{equation}
\noindent
since the quality of service of the consumer depends on the 
total number of base stations deployed by both NSPs.

Finally, with weighted sharing, 
the surplus of a consumer of type $\omega$ subscribing 
to NSP $i$ is
\begin{equation}
 u\Big(\omega, \psi_i, \sum_{k \in \{1,2\}} n_k, p_i \Big) = \omega \mu \psi_i \sum_{k \in \{1,2\}} n_k - p_i
\end{equation}
\noindent
where the utility now also depends on the weight assigned to NSP $i$ in the scheduler, $\psi_i$. 
(We see from Figure~\ref{fig:user-tput} that the average user throughput for a subscriber of NSP $i$
scales linearly with $\psi_i$.)

\subsection{Solution}



We can solve the simple game described above for the best response of each player, 
and use this to gain some insight into the market. To do so, we
define two marginal consumers:

\begin{itemize}
  \item $\underline{\omega}$ is the consumer who is indifferent between subscribing to 
  the smaller NSP (NSP~2) or to neither.
  \item $\overline{\omega}$ is the consumer who is indifferent between subscribing to 
  NSP~1 or NSP~2.
\end{itemize}
We also note that the market share of NSP~1 is
\begin{equation}
\frac{\hat{\omega}-\overline{\omega}}{\hat{\omega}}
\end{equation}
and the market share of NSP~2 is
\begin{equation}
\frac{\overline{\omega}-\underline{\omega}}{\hat{\omega}}.
\end{equation}

Since
\begin{equation}
\overline{\omega}\mu n_1 - p_1 = \overline{\omega}\mu n_2 - p_2
\end{equation}
and 
\begin{equation}
\underline{\omega}\mu n_2 - p_2 = 0
\end{equation}
when there is no resource sharing, we find that
\begin{align}
\underline{\omega} = \frac{p_{2}}{\mu n_{2}}
\label{eq:no-under}
\end{align}
and
\begin{equation}
\overline{\omega} = \frac{p_{1} - p_{2}}{\mu (n_{1} - n_{2})}
\label{eq:no-over}
\end{equation}
and that when $\frac{p_1}{p_2} > \frac{n_1}{n_2}$, there are positive values
of $p_1$, $p_2$, $n_1$ and $n_2$, 
where $0 \leq \underline{\omega} \leq \overline{\omega} \leq \hat{\omega}$
(i.e. both NSPs will have positive market share). 

Substituting (\ref{eq:no-under}) and (\ref{eq:no-over}) into (\ref{eq:nspprofit}) 
and using backward induction, we find that the 
best responses of NSP~1 and NSP~2 are
\begin{align}
p_{1,\text{NS}}^{*} = \frac{{\left(2 \, c_{1} + c_{2}\right)} n_{1} - c_{1} n_{2} + 2 \, \mu\hat{\omega}n_{1}{\left(n_{1}- n_{2}\right)} }{2 \, {\left(2 \, n_{1} - n_{2}\right)}}
\end{align}
and
\begin{align}
&p_{2,\text{NS}}^* = \frac{4  c_{2} n_{1}^{2} + {\big(2 c_{1} - c_{2}+2\mu\hat{\omega}(n_1-n_2)\big)} n_{1} n_{2} - c_{1} n_{2}^{2} }{4 n_{1} {\left(2  n_{1} -  n_{2}\right)}}
\end{align}

In the equal sharing scenario, the consumer selects one NSP over the other
based only on price, since it will get the benefit of all base stations
by subscribing to either NSP. Therefore, the NSPs will have to compete on price
in order to attract subscribers, and if either NSP sets a lower price, 
it will capture all of the subscribers. If both NSPs have the same marginal cost
($c_1 = c_2 = c$), then the best response is
\begin{equation}
p_{1,\text{ES}}^* = p_{2,\text{ES}}^* = c
\end{equation}
and from~(\ref{eq:nspprofit}) we can see that 
both NSPs will earn zero profit. If one NSP has a lower marginal cost, it 
will undercut the other (e.g. if $c_1 < c_2$, then $p_{1}^* = c_2 - \epsilon$)
and capture all of the potential subscribers, leaving the NSP with higher 
marginal cost to have zero market share.
Thus for this simple duopoly game, there is no way for both NSPs to earn a profit
greater than zero with equal sharing.

Finally, we consider the scenario with weighted sharing.
Here, since
\begin{equation}
\overline{\omega}\mu \psi_1 (n_1 + n_2) - p_1 = \overline{\omega}\mu \psi_2 (n_1 + n_2) - p_2
\end{equation}
and 
\begin{equation}
\underline{\omega}\mu \psi_2 (n_1 + n_2) - p_2 = 0
\end{equation}
we find that
\begin{equation}
\underline{\omega} =  \frac{p_{2}}{\mu\psi_{2}{\left( n_{1} + n_{2}\right)} }
\label{eq:ws-under}
\end{equation}
\begin{align}
\overline{\omega} = \frac{p_{1} - p_{2}}{\mu {\left(n_{1} + n_{2}\right)} (\psi_{1} -\psi_{2})}
\label{eq:ws-over}
\end{align}
and that when $\frac{p_1}{p_2} > \frac{\psi_1}{\psi_2}$ and $\psi_1 \geq \psi_2$, there are positive values
of $p_1$, $p_2$, $\psi_1$, $\psi_2$, $n_1$ and $n_2$, 
where $0 \leq \underline{\omega} \leq \overline{\omega} \leq \hat{\omega}$
(i.e. both NSPs will have positive market share). 

Then, substituting (\ref{eq:ws-under}) and (\ref{eq:ws-over}) into (\ref{eq:nspprofit}) 
and using backward induction, we find that the 
the best response of 
NSP~1 is
\begin{align}
p_{1,\text{WS}}^* = \frac{2\mu\hat{\omega}\psi_1(\psi_1-\psi_2)(n_1+n_2)+(2c_1+c_2)\psi_1-c_1\psi_2}{2(2\psi_1-\psi_2)}
\end{align}
and the best response of NSP~2 is
\begin{align}
    \resizebox{0.49\textwidth}{!}{
$p_{2,\text{WS}}^* =\frac{\psi_1\psi_2\big(2\mu\hat{\omega}(n_1+n_2)(\psi_1-\psi_2)+2c_1-c_2\big)+4c_2\psi_1^2-c_1\psi_2^2}{4\psi_1(2\psi_1-\psi_2)}$
}
\end{align}

\subsection{Simplified game with zero marginal costs}


When $c_1 = c_2 = 0$, we show that there are conditions under which 
weighted sharing may be mutually beneficial for both NSPs.
(Recall this in this game, there are no conditions under which 
both NSPs can earn profit greater than zero with \emph{equal} sharing.)

With zero marginal costs, and when there is no resource sharing, NSP~1 earns profit
\begin{equation}
\pi_{1,\text{NS}} = \frac{{\mu \hat{\omega} n_{1}}(n_{1} - n_{2})} {2 \, {\left(2 \, n_{1} - n_{2}\right)}}
\end{equation}
and NSP~2 earns profit
\begin{align}
\pi_{2,\text{NS}} = \frac{{\mu \hat{\omega} n_{1}n_{2} (n_{1}-n_{2})}  }{4 \, {(2n_{1}-n_{2})^2}}
\end{align}

With weighted sharing, NSP~1 earns profit
\begin{align}
\pi_{1,\text{WS}} = \frac{{\mu \hat{\omega} \psi_{1}(\psi_{1}-\psi_{2})\left( n_{1} + n_{2}\right)} }{2 \, {\left(2 \, \psi_{1} - \psi_{2}\right)}}
\end{align}
and NSP~2 earns profit
\begin{align}
\pi_{2,\text{WS}} = \frac{\mu \hat{\omega} \psi_{1} \psi_{2}(\psi_{1}-\psi_{2}){\left( n_{1} + n_{2}\right)} }{4 \, {\left(2 \, \psi_{1}- \psi_{2}\right)^2}}
\end{align}

We are interested in finding values of $n_1$, $n_2$, $\psi_1$, and $\psi_2=1-\psi_1$ for which
weighted sharing is more profitable for both NSPs than no sharing, i.e.
$\pi_{1,\text{WS}} > \pi_{1,\text{NS}}$ and $\pi_{2,\text{WS}} > \pi_{2, \text{NS}}$.
For NSP~1, we find that if $\psi_1 > \frac{n_1}{n_1+n_2}$
then $\pi_{1,\text{WS}} > \pi_{1,\text{NS}}$.
For NSP~2, if there exists values for 
$n_1$, $n_2$, $\psi_1$, and $\psi_2=1-\psi_1$
such that
\begin{align}
\frac{n_1}{n_1+n_2} < \psi_1 <  \frac{4 \, n_{1}^{2} - 5 \, n_{1} n_{2} + 3 \, n_{2}^{2} + \sqrt{\Delta}}{4 \, {\left(2 \, n_{1}-n_{2}\right)^{2}}},
\end{align}
with $\Delta=16n_{1}^{4} - 8 n_{1}^{3} n_{2} - 15 n_{1}^{2} n_{2}^{2} + 10 n_{1} n_{2}^{3} + n_{2}^{4}$, then $\pi_{2,\text{WS}} > \pi_{2, \text{NS}}$
(and under these conditions, we also have $\pi_{1,\text{WS}} > \pi_{1,\text{NS}}$, so 
weighted sharing is mutually beneficial).

\begin{figure}
\centering
\includegraphics[width=3in]{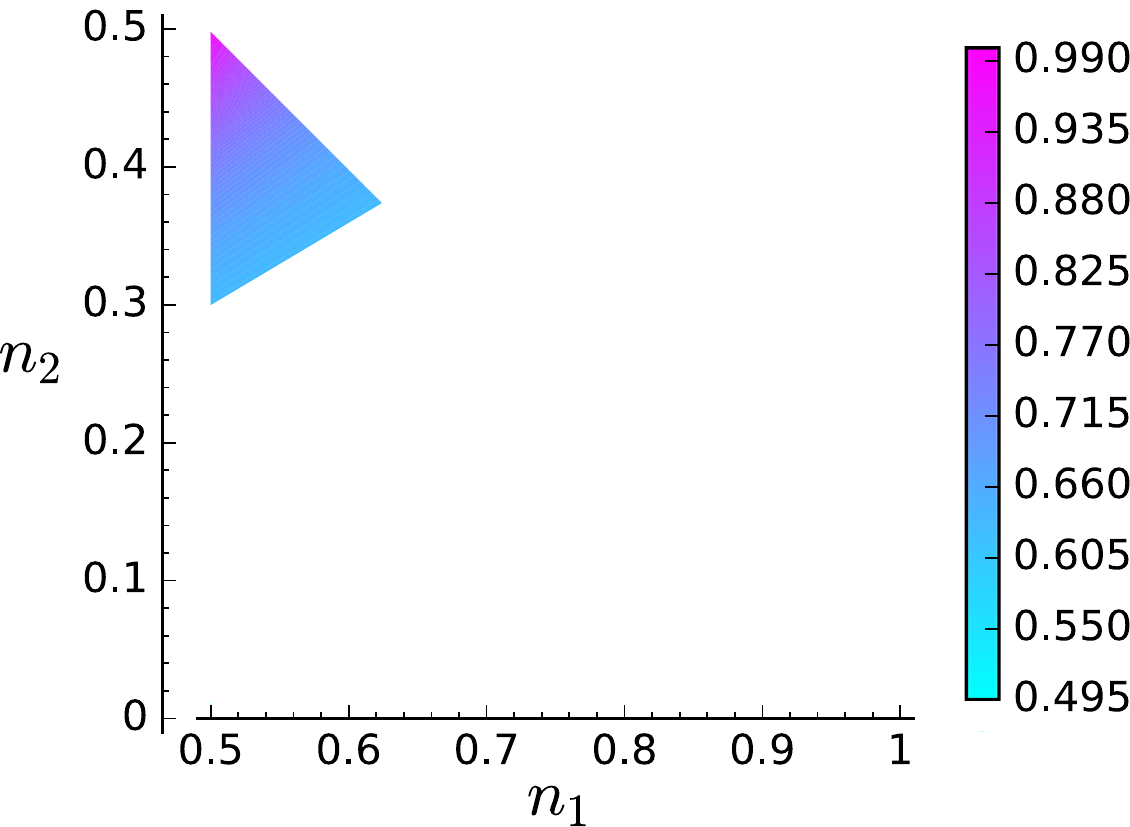}
\caption[]{The shaded region shows the values of $n_1$ and $n_2$ ($n_1 > n_2$) where there is a positive $\psi_1$ and $\psi_2 = 1 - \psi_1$ such that
$\pi_{1,\text{WS}} > \pi_{1,\text{NS}}$, $\pi_{2,\text{WS}} > \pi_{2,\text{NS}}$, and $n_1 + n_2 \leq 1$. Furthermore, the depth of the shading
gives the maximum value of $\psi_1$ for which this holds. The minimum value of $\psi_1$ for which sharing 
is profitable for both NSPs is $\frac{n_1}{n_1 + n2}$.}
 \label{fig:region}
\end{figure}

Figure~\ref{fig:region} illustrates the range of $n_1$ and $n_2$ in which sharing can be mutually beneficial, i.e., where there is a positive $\psi_1$ and $\psi_2 = 1 - \psi_1$ such that
$\pi_{1,\text{WS}} > \pi_{1,\text{NS}}$, $\pi_{2,\text{WS}} > \pi_{2,\text{NS}}$, and $n_1 + n_2 \leq 1$. 
In general, weighted sharing is most helpful when NSPs are similar
in size. Ordinarily, similar-sized NSPs must compete on price,
since there is little else to distinguish them; in a weighted sharing
scenario, however, they can vary $\psi_1$ in order to differentiate
themselves and make price competition less tough. 
For NSPs that are already dissimilar, the benefit 
of weighted sharing for easing price competition 
does not apply.

Next, we discuss the specific values of $\psi_1$ and 
$\psi_2 = 1 - \psi_1$ for which sharing is mutually beneficial.
For weighted sharing to be more profitable than no sharing 
for NSP~1, $\psi_1$
should be at least $\frac{n_1}{n_1 + n2}$, i.e., the share of the 
airtime allocated to NSP~1 should be at least equal to its share 
of the resources. The maximum value of $\psi_1$ for which 
sharing is mutually beneficial is shown in Figure~\ref{fig:region},
and may be somewhat larger than $\frac{n_1}{n_1 + n2}$, 
especially when $n_1$ is not much larger than $n_2$. 
The intuition behind this surprising result is based on the dynamics
of price competition between the NSPs.
The more similar the sizes of the NSPs, the tougher the price 
competition gets, as consumers differentiate between 
them mainly based on price. Thus with no sharing, similar-sized NSPs
must set a low price and make little profit.
With weighted sharing and large $\psi_1$,
however, the NSPs are differentiated in the quality of service
they offer to consumers. Subject to less price competition, both
NSPs can set a higher price and earn more profit than they would
with no sharing, 
even though the quality of service
offered by NSP~2 is lower with weighted sharing and large $\psi_1$
than with no sharing at all, as shown in Figure~\ref{fig:user-tput}.

Similar-sized NSPs could also be differentiated in quality 
by making $\psi_1$ small, so that the quality of NSP~2's service 
would be much better than NSP~1's. However, because of the greater 
market power afforded to NSP~1 by its position in the game, it 
can generally set higher prices without sharing and so it would not 
be more profitable with weighted sharing and 
$\psi_1 < \frac{n_1}{n_1 + n2}$.

The key result of this section is that \emph{there exists} a range of 
market conditions ($n_1$ and $n_2$) in the duopoly game, 
for which weighted sharing can be more profitable than no sharing 
for both NSPs. Furthermore, in Section~\ref{sec:simulation}, we have shown 
that weighted sharing achieves almost the full sharing gains
from a purely technical perspective (in terms of network capacity). 
Thus, under the right market conditions, base station sharing with a 
weighted scheduler may be beneficial from both a technical and economic perspective.

\section{Conclusions}
\label{sec:conclusion}

While resource sharing, and base station sharing in particular, 
can increase overall network capacity in a mmWave cellular system,
NSPs may still be unwilling to share due to unfavorable
competitive dynamics with equal sharing. In this work,
we describe a scheduling approach with which we may:
\begin{itemize}
\item achieve an overall network capacity similar to 
an equal sharing scenario, capturing most of the potential gains
due to BS diversity and multiuser diversity, and
\item still maintain enough of a competitive difference 
between asymmetric NSPs in a duopoly market so that sharing is profitable for 
both.
\end{itemize}
In particular, with this approach, an NSP can achieve technical sharing gains without
losing its competitive edge in the market.

We briefly discuss here some limitations of our approach.
We make some approximations for tractability, such as 
approximating the average data rate as a linear function 
of the number of base stations, and approximating the 
average number of users in a cell as fixed in Section~\ref{sec:game}
(when in fact this depends on the consumers' decisions in the game).
We also use a very simple model for the game described 
in Section~\ref{sec:game}, although we believe this model
sufficiently captures the key details of the market 
to support our general conclusions.

As future work, we would like to extend the model 
of the duopoly game in Section~\ref{sec:game} to model 
the dependence of consumers' decisions on 
the network data rate. We would also like to consider
the application of our weighted sharing 
scheduler to other kinds of networks, such as small 
cell microwave networks and WiFi networks.

\section*{Acknowledgment}

This work was supported by the National Science Foundation
under Grant No. 1547332, 1302336, and the Graduate Research Fellowship Program, 
by the New York State Center for Advanced Technology in
Telecommunications (CATT),
and by NYU WIRELESS.



%
\bibliographystyle{IEEEtran}
\bibliography{mmSys}

\begin{thebibliography}{10}
\providecommand{\url}[1]{#1}
\csname url@samestyle\endcsname
\providecommand{\newblock}{\relax}
\providecommand{\bibinfo}[2]{#2}
\providecommand{\BIBentrySTDinterwordspacing}{\spaceskip=0pt\relax}
\providecommand{\BIBentryALTinterwordstretchfactor}{4}
\providecommand{\BIBentryALTinterwordspacing}{\spaceskip=\fontdimen2\font plus
\BIBentryALTinterwordstretchfactor\fontdimen3\font minus
  \fontdimen4\font\relax}
\providecommand{\BIBforeignlanguage}[2]{{%
\expandafter\ifx\csname l@#1\endcsname\relax
\typeout{** WARNING: IEEEtran.bst: No hyphenation pattern has been}%
\typeout{** loaded for the language `#1'. Using the pattern for}%
\typeout{** the default language instead.}%
\else
\language=\csname l@#1\endcsname
\fi
#2}}
\providecommand{\BIBdecl}{\relax}
\BIBdecl

\bibitem{andrews-sharing}
A.~K. Gupta, J.~G. Andrews, and R.~W. Heath, ``On the feasibility of sharing
  spectrum licenses in {mmWave} cellular systems,'' \emph{IEEE Transactions on
  Communications}, vol.~64, no.~9, pp. 3981--3995, Sept 2016.

\bibitem{matia}
M.~Rebato, M.~Mezzavilla, S.~Rangan, and M.~Zorzi, ``Resource sharing in {5G
  mmWave} cellular networks,'' in \emph{2016 IEEE Conference on Computer
  Communications Workshops (INFOCOM WKSHPS)}, April 2016, pp. 271--276.

\bibitem{european-sharing}
F.~Boccardi, H.~Shokri-Ghadikolaei, G.~Fodor, E.~Erkip, C.~Fischione,
  M.~Kountouris, P.~Popovski, and M.~Zorzi, ``Spectrum pooling in mmwave
  networks: Opportunities, challenges, and enablers,'' \emph{IEEE
  Communications Magazine}, vol.~54, no.~11, November 2016.

\bibitem{andrews-sharing-second}
A.~K. Gupta, A.~Alkhateeb, J.~G. Andrews, and R.~W. Heath, ``Gains of
  restricted secondary licensing in millimeter wave cellular systems,''
  \emph{IEEE Journal on Selected Areas in Communications}, vol.~34, no.~11, pp.
  2935--2950, Nov 2016.

\bibitem{fund2017resource}
F.~Fund, S.~Shahsavari, S.~S. Panwar, E.~Erkip, and S.~Rangan, ``Resource
  sharing among mmwave cellular service providers in a vertically
  differentiated duopoly,'' \emph{IEEE International Conference on
  Communications (ICC)}, 2017.

\bibitem{singhProfitSharing}
C.~Singh, S.~Sarkar, A.~Aram, and A.~Kumar, ``Cooperative profit sharing in
  coalition-based resource allocation in wireless networks,'' \emph{IEEE/ACM
  Transactions on Networking}, vol.~20, no.~1, pp. 69--83, Feb. 2012.

\bibitem{Iturralde2013}
M.~Iturralde, A.~Wei, T.~Ali-Yahiya, and A.-L. Beylot, ``Resource allocation
  for real time services in lte networks: Resource allocation using cooperative
  game theory and virtual token mechanism,'' \emph{Wireless Personal
  Communications}, vol.~72, no.~2, pp. 1415--1435, Sep 2013.

\bibitem{allocateFairPayoff}
J.~Cai and U.~Pooch, ``Allocate fair payoff for cooperation in wireless ad hoc
  networks using {Shapley} value,'' in \emph{18th International Parallel and
  Distributed Processing Symposium}, April 2004.

\bibitem{eyeball}
R.~T.~B. Ma, D.~M. Chiu, J.~C.~S. Lui, V.~Misra, and D.~Rubenstein, ``On
  cooperative settlement between content, transit, and eyeball internet service
  providers,'' \emph{IEEE/ACM Transactions on Networking}, vol.~19, no.~3, pp.
  802--815, June 2011.

\bibitem{shahram-scheduler}
S.~Shahsavari and N.~Akar, ``A two-level temporal fair scheduler for multi-cell
  wireless networks,'' \emph{IEEE Wireless Communications Letters}, vol.~4,
  no.~3, pp. 269--272, June 2015.

\bibitem{mustafa-channel}
M.~R. Akdeniz, Y.~Liu, M.~K. Samimi, S.~Sun, S.~Rangan, T.~S. Rappaport, and
  E.~Erkip, ``Millimeter wave channel modeling and cellular capacity
  evaluation,'' \emph{IEEE Journal on Selected Areas in Communications},
  vol.~32, no.~6, pp. 1164--1179, 2014.

\bibitem{heath-mm}
T.~Bai and R.~W. Heath, ``Coverage and rate analysis for millimeter-wave
  cellular networks,'' \emph{IEEE Transactions on Wireless Communications},
  vol.~14, no.~2, pp. 1100--1114, 2015.

\end{thebibliography}

\end{document}